\documentclass[runningheads]{svmult}

\usepackage{makeidx}   
\usepackage{graphicx}  
\usepackage{subeqnar}  
\usepackage{multicol}  
\usepackage{physprbb}  
\makeindex             

\begin{document}
\title*{Globular Clusters in nearby Galaxy Clusters}
\toctitle{Globular Clusters in nearby Galaxy Clusters}
\titlerunning{Globular Clusters in nearby Galaxy Clusters}
\author{Michael Hilker}
\authorrunning{Michael Hilker}
\institute{Sternwarte der Universit\"at Bonn, Auf dem H\"ugel 71, 53121 Bonn,
Germany}

\maketitle              

\begin{abstract}
The discovery of a large population of intra-cluster
star clusters in the central region of the Hydra~I and Centaurus 
galaxy clusters is presented. Based on deep VLT photometry ($V,I$), many star 
clusters have been identified not only around the early-type galaxies, but also 
in the intra-cluster field, as far as 250 kpc from the galaxy centres.
These intra-cluster globulars are predominantly blue, with a siginificant 
fraction 
being even bluer than the metal-poor halo clusters around massive galaxies.
When interpreted as a metallicity effect they would have iron abundances
around [Fe/H]$\simeq-2$ dex. However, they might also be relatively young
clusters which could have formed in tidal tails during recent interactions of 
the central galaxies.
\end{abstract}

\section{Introduction}
In galaxy clusters the globular cluster (GC) population of a large variety of 
galaxies, ranging from giant ellipticals to dwarf galaxies, can be studied.
As it turns out, the results of these studies might be very different from 
that of isolated galaxies in the field.

The most striking discovery in clusters is that the central giant galaxy
(mostly a cD elliptical) often possesses an extra-ordinary rich globular
cluster system ($\simeq$6000-10000 GCs) \cite{harr91} and also a high specific 
frequency ($S_N\simeq6$-10) as compared to ``normal'' ellipticals 
($S_N\simeq3$).

To explain these findings various formation scenarios have been proposed. 
Among them are 1) the merging of major galaxies \cite{ashm92}, 2) the accretion 
of large numbers of dwarf galaxies/small fragments
\cite{hilk99},\cite{cote02}, or 3) a large, multiple dissipational collapse 
without a significant fraction of accretion or merger events \cite{forb97}.
One might here object that the true picture probably is very complex and that
different mechanisms have been at work and acted at different epochs with
changing efficiency. A first step towards a more realistic picture probably is
to follow the formation of GC populations within the framework of a 
cosmologically motivated hierarchical merger tree of dark matter halos 
\cite{beas02}.

Cosmological simulations for the formation and evolution of galaxy clusters
show that the massive central galaxies are built up by continuous
accretion of several smaller galaxies, especially in the first Giga years
of structure formation \cite{dubi98}. However, also presently
small groups of galaxies fall into the centres of clusters.
It is natural to ask what happened with the globular clusters of the merged
and accreted galaxies. How are they distributed today {\em ``after all the 
mess''}? Have new clusters perhaps been formed in the interactions?
Nearby galaxy clusters provide an ideal laboratory to study these issues
in detail.

\section{Globular cluster photometry in nearby galaxy clusters}

In the following a galaxy cluster shall be defined as `nearby' if it fulfils
two conditions: 1) globular clusters are resolved as point sources, and 2)
photometry can reach the turnover of the globular cluster luminosity function.
Obviously the distance of galaxy clusters where these constraints are
valid depends on the telescope one uses and the observing conditions.

\begin{figure}[t]
\begin{center}
\includegraphics[width=.8\textwidth]{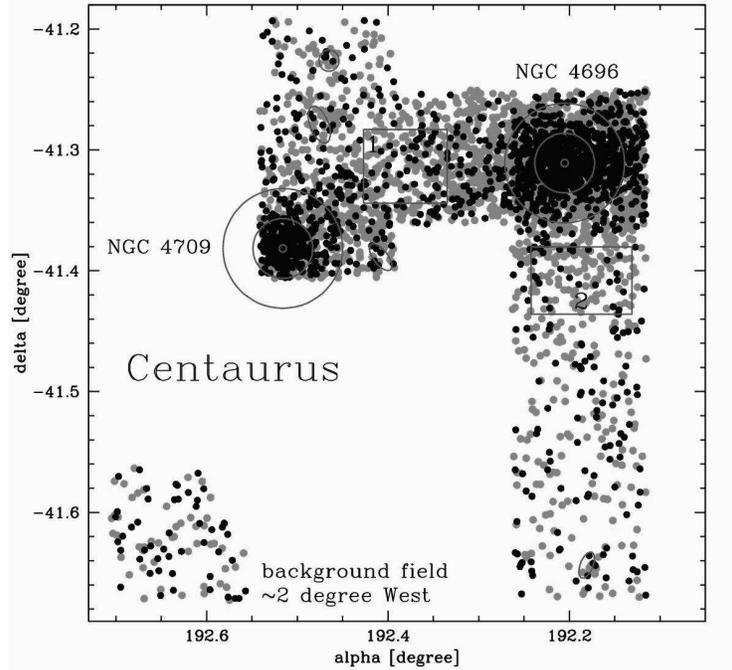}
\end{center}
\caption[]{Projected spatial distribution of blue ($0.7<(V-I)<1.0$, grey dots) 
and red ($1.0<(V-I)<1.3$, dark dots) globular cluster candidates in the
centre of the Centaurus cluster is shown. Ellipses indicate the location
of the major galaxies. Circles and boxes mark the selected regions for the
colour distribution of globular clusters in Fig.~\ref{col}.}
\label{centaur}
\end{figure}

Reachable clusters for 4m-class telescopes are the ones within a distance
of about 30 Mpc (Virgo, Fornax and Antlia). Wide-field imagers can be used to
cover their core area ($r_{\rm core}\simeq150$ kpc). Indeed many observations
are underway to study the GCSs in these clusters, e.g. Dirsch et al. (this
volume).

The GCs in galaxy clusters as far as 50 Mpc can be imaged by 8m-class
telescopes. The HST might even reach further out ($\simeq80$ Mpc, e.g. Coma
cluster), but its usefulness is restricted by its small field-of-view. In this 
contribution, 
we present the photometric mapping of the core regions of the Hydra~I and 
Centaurus galaxy clusters. Both are located at a distance of about 45 Mpc.

Hydra~I and Centaurus were observed at dark time and under photometric
conditions in the filters $V$ and $I$ with FORS1 at the VLT (ESO, Paranal).
The seeing was in the range 0.5 to 0.7 arcsec, providing well resolved and
very homogeneous data. Fig.~\ref{centaur} shows the seven pointings in 
Centaurus.

The Centaurus cluster is dynamically young with two merging sub-groups, a
main cluster component (Cen30) around the cD galaxy NGC~4696 and a smaller
group component (Cen45) around NGC~4709 \cite{stei97}.
The Hydra~I cluster is dynamically evolved, has a regular core shape
and an isothermal X-ray gas halo \cite{tamu00}.
However, a small group of late-type galaxies around the spiral NGC~3312 just 
seems to cruise through the cluster centre.

\begin{figure}[t]
\begin{center}
\includegraphics[width=1.0\textwidth]{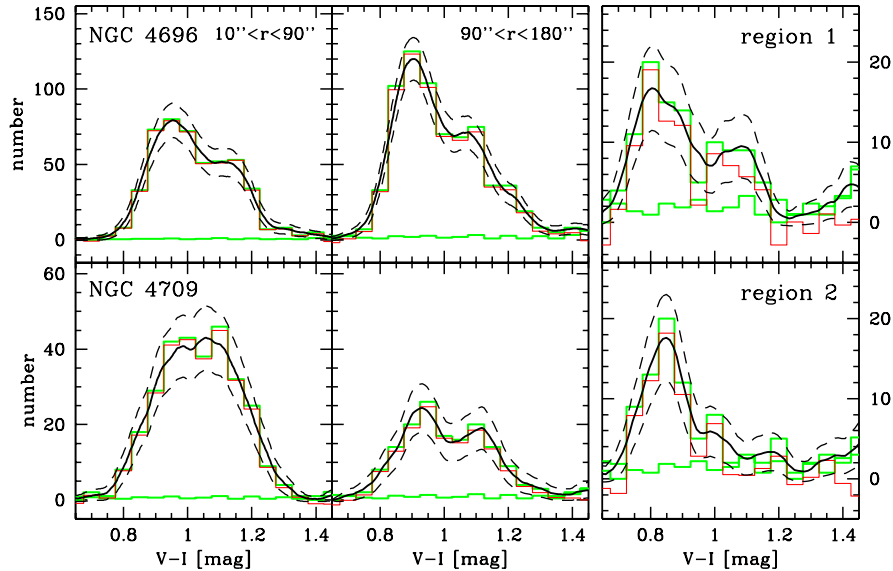}
\end{center}
\caption[]{Colour distribution of globular cluster candidates around NGC~4696
(upper left panels), around NGC~4709 (lower left panels), and the two
intra-cluster regions from Fig.~1 (right panels). The histograms are the
background counts, the raw counts nad the background corrected counts. The 
curves are binning independent representations of the number counts and their 
1$\sigma$ uncertainty limits (dashed curves).}
\label{col}
\end{figure}

\section{Results for Hydra~I and Centaurus}
\begin{figure}[t]
\begin{center}
\includegraphics[width=1.0\textwidth]{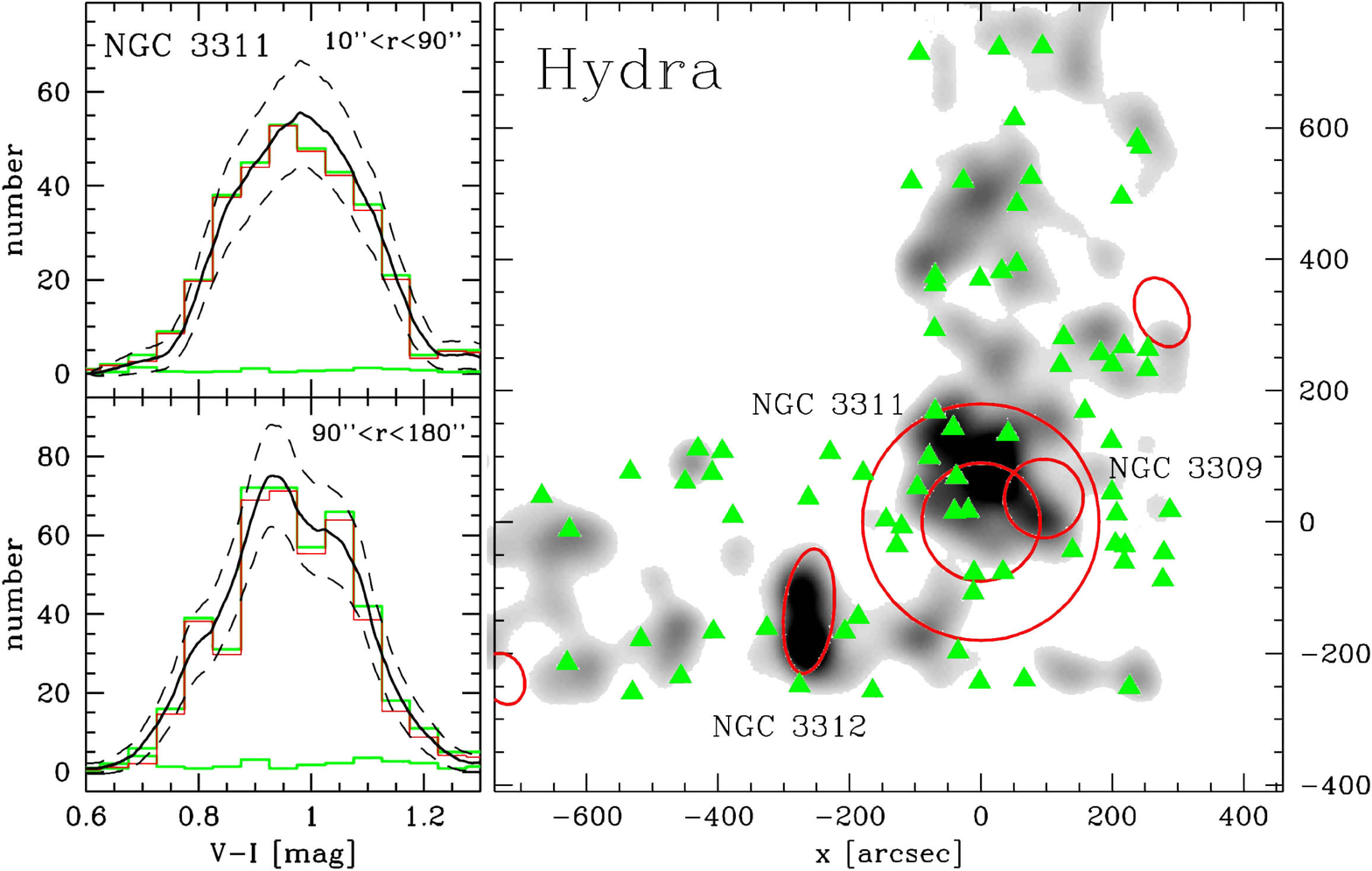}
\end{center}
\caption[]{Left panel: colour distribution of cluster candidates around
NGC~3311 in two rings as indicated at right. The curves have the same
meaning as in Fig.~\ref{col}. Right panel: smoothed density map of the very
blue cluster candidates ($0.5<(V-I)<0.8$). The ellipses indicate the positions
of the major galaxies. Grey triangles mark newly discovered dwarf galaxies.}
\label{hydra}
\end{figure}

In both galaxy clusters, many star clusters have been identified down to the
turnover magnitude of the globular cluster luminosity function at 
$V\simeq26.0$ mag. They are distributed not only around the early-type 
galaxies, but also in the intra-cluster field, as far as 250 kpc from the 
cluster centres (see Figs.~\ref{centaur} and \ref{hydra}). This is well
outside the tidal radii of the central galaxies. 

\begin{figure}[t]
\begin{center}
\includegraphics[width=1.0\textwidth]{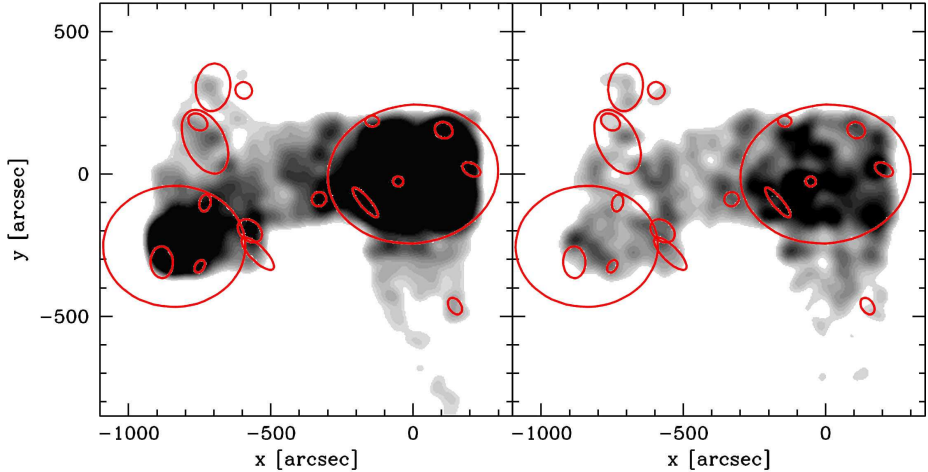}
\end{center}
\caption[]{Smoothed density maps of all globular cluster candidates (left) and
very blue ($0.7<(V-I)<0.85$) cluster candidates (right) are shown.
The ellipses indicate the faintest measurable isophotes of the major galaxies 
at a surface brightness of about $\mu_V=27.8$ mag/arcsec$^2$. There are clear
overdensities of clusters outside these isophotes.}
\label{dens}
\end{figure}

In the Centaurus cluster, the intra-cluster star clusters show a tidal 
tail like structure between the two dominant giant ellipticals (see 
Fig.~\ref{dens}). In Hydra~I, they are distributed asymmetrically
around the central galaxy NGC~3311 and occupy the same space as the abundant
(newly found) dwarf spheroidal galaxies (see Fig.~\ref{hydra}). However, in 
both clusters the spatial coverage of the observations is not sufficient to 
draw conclusions
about their nature. The overdensity of star clusters in the intra-cluster
field is 3-4 times higher than for objects with the same selection criteria in 
a background control field (see Fig.~\ref{dens}). Since the intra-cluster star
light is very faint ($>28$ mag/arcsec$^2$), the estimated specificic frequency 
of the intra-cluster globulars is very high ($S_N>10$).

When looking at the colour distribution of globular clusters around the giant
ellipticals in Centaurus and Hydra~I (Fig.~\ref{col}), one first recognizes 
the familiar bimodality, with peaks around
$(V-I)\simeq0.9$ and $(V-I)\simeq1.1$ mag. The bimodality is more pronounced
in the outer parts of the galaxies, see Fig.~\ref{col} (middle panels) and
Fig.~\ref{hydra} (lower left panel). Also the ratio of blue to red clusters
increases with galacto-centric distance. Already in Fig.~\ref{centaur} one
notices that blue clusters are more widely distributed than the red ones, which
are concentrated towards the galactic bulges.

The colour distributions of the intra-cluster star clusters in Centaurus is
dominated by blue clusters, as expected. But surprisingly, they peak around
$(V-I)\simeq0.8$ mag (see Fig.~\ref{col}, right panels) which is slightly but
noticeably bluer than the old metal-poor clusters in the halos of the giant
galaxies. Also in the outer halo of NGC~3311 in Hydra~I clusters with very
blue colours are found (see Fig.~\ref{hydra}). These clusters might have been
stripped from the late-type group of galaxies around NGC~3312 that is passing
by the core of Hydra~I.

\section{Interpretation of the very blue cluster population}
The blue colour of the intra-cluster clusters can be interpreted in two ways. 
Either they are old and even more metal-poor than the metal-poor halo clusters 
in massive galaxies, or their blue colour indicates a young age and presumably 
higher metallicity. How would these clusters, far in the outskirts of the 
bulges of the central galaxies, fit into the picture of the different formation
scenarios as presented in Sect.~1?

{\bf First} let us suppose that these clusters are {\bf old and metal-poor}. 
According to their $(V-I)$ colours their iron abundance would be around $-2.0$ 
dex \cite{kiss98a}. Very metal-poor globular clusters
are known to exist around low mass dwarf ellipticals (e.g. Lotz et al., this
volume), or might have
formed in small proto-galactic fragments that were incorporated into the halo
of a larger galaxy
before becoming dwarf galaxies themselves. The large extended halo around 
central cluster galaxies might then have formed via the accretion of a large
number of dwarf galaxies whose clusters have been stripped during their infall.
In this case one would expect that the metal-poor GCs are distributed quite 
uniformly around the central galaxy, since any assymmetric accretion of 
metal-poor dwarfs/fragments at the formation epoch of the halo should have 
vanished over a Hubble time.

{\bf Second} let us suppose that the blue clusters are {\bf young and rather
metal-rich}. Several age/metallicity combinations seem possible according
to their $(V-I)$ colour only. But how could young clusters
exist that far outside the bulges of their host galaxies without any 
sign of a corresponding young field star population and gas in their vicinity? 
One might imagine that these clusters have formed from material that has been
thrown out in large tidal tails during a major merger. All the gas would have 
been consumed 
in this event and/or some rest gas was stripped rapidly and would fall into 
the centre of the merger remnant. The timescales are short, since any material
could travel hundreds of kpc in much less than a Gyr, when assuming a 
reasonable velocity of few hundreds km/sec.
Interestingly, in the Centaurus as well as in the Hydra~I cluster, the central
galaxy (NGC~4696 and NGC~3311) possesses prominent dust lanes in its centre
that point to a rather recent accretion or merging event.

The feasability of both scenarios makes a further investigation of the 
intra-cluster stellar population in both clusters very interesting and urgent.

\section{What still has to be done}

The presented data set still hasn't been exploited fully. There are several 
issues in globular cluster astronomy that haven't been touched yet by our
analyses. As next steps we intend to 1) study the specific frequency 
of intra-cluster GCs as a function of centro-cluster distance, 2) study the
individual GCSs for all member galaxies down to the dwarf galaxy regime 
(colours, density profiles, $S_N$, etc.), and 3) correlate the properties of 
the central GCS with those of the cD halo, the dwarf galaxy population and 
the X-ray 
gas. Moreover, follow-up $U$ band photometry and multi object spectroscopy
are planned to break the age-metallicity degeneracy for the blue clusters and
confirm the membership of bright GC candidates and dwarf galaxies.

%

\end{document}